\documentclass[journal]{IEEEtran}
\IEEEoverridecommandlockouts

\usepackage{cite}
\usepackage{graphics} 
\usepackage{epsfig} 
\usepackage{mathptmx} 
\usepackage{times} 
\usepackage{amsmath} 
\usepackage{amssymb}  
\usepackage{multirow}
\usepackage{booktabs}
\usepackage{fourier-orns}
\usepackage{setspace}
\usepackage{graphicx}
\usepackage[justification=centering]{caption}
\usepackage{float}
\usepackage[numbers]{natbib}
\usepackage{verbatim}
\usepackage{algorithmic}
\DeclareMathOperator*{\argmax}{argmax}
\DeclareMathOperator*{\argmin}{argmin}
\usepackage[ruled]{algorithm2e}

\begin{document}
	
\title{A SUMO Framework for Deep Reinforcement Learning Experiments Solving Electric Vehicle Charging Dispatching Problem}

\author{Yaofeng~Song,
	    Han~Zhao,~\IEEEmembership{Student Member,~IEEE,}
     	Ruikang~Luo,~\IEEEmembership{Student Member,~IEEE,}
	    Liping~Huang,~\IEEEmembership{Member,~IEEE,}
	    Yicheng~Zhang,~\IEEEmembership{Member,~IEEE,}
        and~Rong~Su,~\IEEEmembership{Senior Member,~IEEE}

\thanks{*Yaofeng Song, Han Zhao and Ruikang Luo contributed equally to this work and should be considered as joint first authors}%
\thanks{Yaofeng Song is affiliated with School of Electrical and Electronic Engineering, Nanyang Technological University, 639798, Singapore Email: song0223@e.ntu.edu.sg@e.ntu.edu.sg}
\thanks{Han Zhao is affiliated with School of Electrical and Electronic Engineering, Nanyang Technological University, 639798, Singapore Email: ZHAO0278@e.ntu.edu.sg}
\thanks{Ruikang Luo is affiliated with Continental-NTU Corporate Lab, Nanyang Technological University, 50 Nanyang Avenue, 639798, Singapore Email: ruikang001@e.ntu.edu.sg}
\thanks{Liping Huang is affiliated with School of Electrical and Electronic Engineering, Nanyang Technological University, 639798, Singapore Email: liping.huang@ntu.edu.sg}
\thanks{Yicheng Zhang is affiliated with Institute for Infocomm Research (I2R), Agency for Science, Technology and Research (ASTAR), 138632, Singapore Email: zhang$\_$yicheng@i2r.a-star.edu.sg}%
\thanks{Rong Su is affiliated with Division of Control and Instrumentation, School of Electrical and Electronic Engineering, Nanyang Technological University, 50 Nanyang Avenue, Singapore 639798. Email: rsu@ntu.edu.sg}
}

\markboth{IEEE TRANSACTIONS ON VEHICULAR TECHNOLOGY}%
{Shell \MakeLowercase{\textit{et al.}}: Bare Demo of IEEEtran.cls for IEEE Journals}

\maketitle

\begin{abstract}
In the modern city, the number of Electric Vehicle (EV) is increasing rapidly for its low emission and better dynamic performance, leading to an increasing demand of EV charging. However, due to the limited number of EV charging facilities, catering the huge demand of the time consuming EV charging becomes a critical problem. It is quite a challenge to dispatch EVs in the dynamic traffic environment and coordinate interaction among agents. To better serve further research on various related Deep Reinforcement Learning (DRL) EV dispatching algorithms, a efficient simulation environment is necessary to ensure the success. As simulator Simulation Urban Mobility (SUMO) is one of the most widely used open-source simulator, it has great significance to create the environment that satisfies research requirements on SUMO. We aim to improve the efficiency of the EV charging station usage and save time for EV users in the further work, therefore design an EV navigation system on the basis of the traffic simulator SUMO using Jurong Area, Singapore in this paper. Various state-of-the-art DRL algorithms are deployed on the designed testbed to validate the feasibility of the framework in terms of EV charging dispatching problem. Besides EV dispatching problem, the environment can also serve for other reinforcement learning (RL) traffic control problems.

\end{abstract}


\IEEEpeerreviewmaketitle

\section{Introduction}
\IEEEPARstart{G}{lobal} warming and earth resource usage have been serious problems that have been bothering the world for a long time. These problems are caused by different factors and transportation is one of them. According to recent research, 22$\%$ of the total CO2 emission is contributed by transportation in 2020\cite{giannakis2020land}. Moreover, the transportation field takes up 30$\%$ of the global energy consumption, consuming a great amount of gasoline and other forms of energy\cite{chen2019global}. In this condition, EV stands out from different kinds of vehicles for their low energy consumption and environmental friendliness. What’s more, the advance in reliable and efficient electric motors, EV control systems and techniques, Grid-to-Vehicle (G2V) and Vehicle-to-Grid (V2G) for example, that support different functions of EV transportation systems enable EVs to become more and more popular in the vehicle market. In 2021, the worldwide EV sales reached about 675 million which is 208$\%$ as the amount in 2020\cite{murdock2021renewables}.

Even though EV techniques are becoming more and more reliable, multiple challenges are still hindering EVs from replacing traditional vehicles completely. First of all, many EV users suffer from the charging demand of their EVs\cite{bae2011spatial}. The distance that an EV can travel is limited and the time it takes for battery charging is much bigger than the refuelling of traditional vehicles. Second, the ability of EV battery degrades which means that, EV users are required to change their battery after a certain period of travelling. Even though EV utilizes cheap and environment friendly energy, those replaced batteries will still result in pollution because of the battery degradation difficulty, and replacing the battery is expensive. Third, the growing number of EVs in urban cities will lead to the growing demand for charging facilities, public or private. Deploying charging points in the parking lots and managing charging behavior could be a challenge to the government\cite{gonzalez2019impact}. Moreover, customer acceptance, EV chips, battery storage and the reliance on rare earth materials of EVs are all factors that impede EVs from grabbing a larger market share.

Nowadays, a large number of researchers are committed to improving the practicality and reliability of EVs\cite{luo2021deep}. Because of their great contribution, more and more people are willing to choose EVs rather than traditional vehicles, alleviating the pressure of pollution and energy consumption. For the power area, there is research about resource allocation, EV charging scheduling, charging strategy design or charging behavior management. In those works, many optimization methods are proposed by researchers which aim to maximize or minimize the objective function such as the profit of the charging station or the amount of energy consumption, while satisfying different constraints. For example, authors in \cite{ding2020optimal} proposed an optimal EV charging strategy with Markov Decision Process (MDP) and RL technique. This method helps maximize the operator profits of the distribution power network while satisfying grid-level constraints. As for the Intelligent Traffic System (ITS), there are papers about EV fleet management, routing EVs to pick up customers and managing the charging behavior of the EV fleet with minimum travel and queuing time. For example, paper \cite{liang2020mobility} presents a shared on-demand EV fleet charging scheduling method using DRL. The proposed method performs well in EV fleet joint charging scheduling, order dispatching for shard on-demand EVs picking up passengers, and vehicle rebalancing for more efficient usage of the whole EV fleet.

Compared with recent related works, the main contributions can be concluded as follows:
\begin{itemize}
	\item A novel EV navigation environment is built using SUMO.
	
	\item The environment is evaluated as suitable for applying different RL methods.
\end{itemize}

The rest content is arranged as follows: The second section describes related research about the intelligent vehicle dispatching system and commonly used optimization algorithms. The third section illustrates the problem statement and detailed EV navigation system design with the SUMO simulation environment setup. The fourth section evaluates the EV navigation environment by applying different deployment methods including mathematical methods and RL approaches. And the final section summarizes the contribution and possible future plan.

\section{RELATED Research}
In this section, related work on the intelligent vehicle dispatching system and related RL techniques on optimization will be introduced.

RL is a useful tool in the ITS design\cite{rasheed2020deep}. Based on the interaction with the environment, RL agent can provide the best solution which could be regarded as the solution that most possibly leads to the greatest reward\cite{wiering2012reinforcement}. The method works as follows, the environment would provide the current state of the object to the RL agent. On the basis of the current state information, the agent calculates the answer that will lead to the most immediate reward and the greatest potential reward in the future. Given the best solution, the agent interacts with the environment, getting the reward of this action and the new state of the controlled object. From numerous times of this cycle, taking action and retrieving feedback, the RL agent learns from multiple times of attempts and become able to provide the most reliable answer on the basis of the policy it concludes. The requirement of hundreds or thousand of attempts and learning from the experience makes RL method a data-driven method that iteratively computes the most promising but approximate solution. Therefore, RL method is also regarded as approximate dynamic programming\cite{sutton2018reinforcement}.
\begin{figure}[!htb]
	\centering
	\includegraphics[width=1\linewidth]{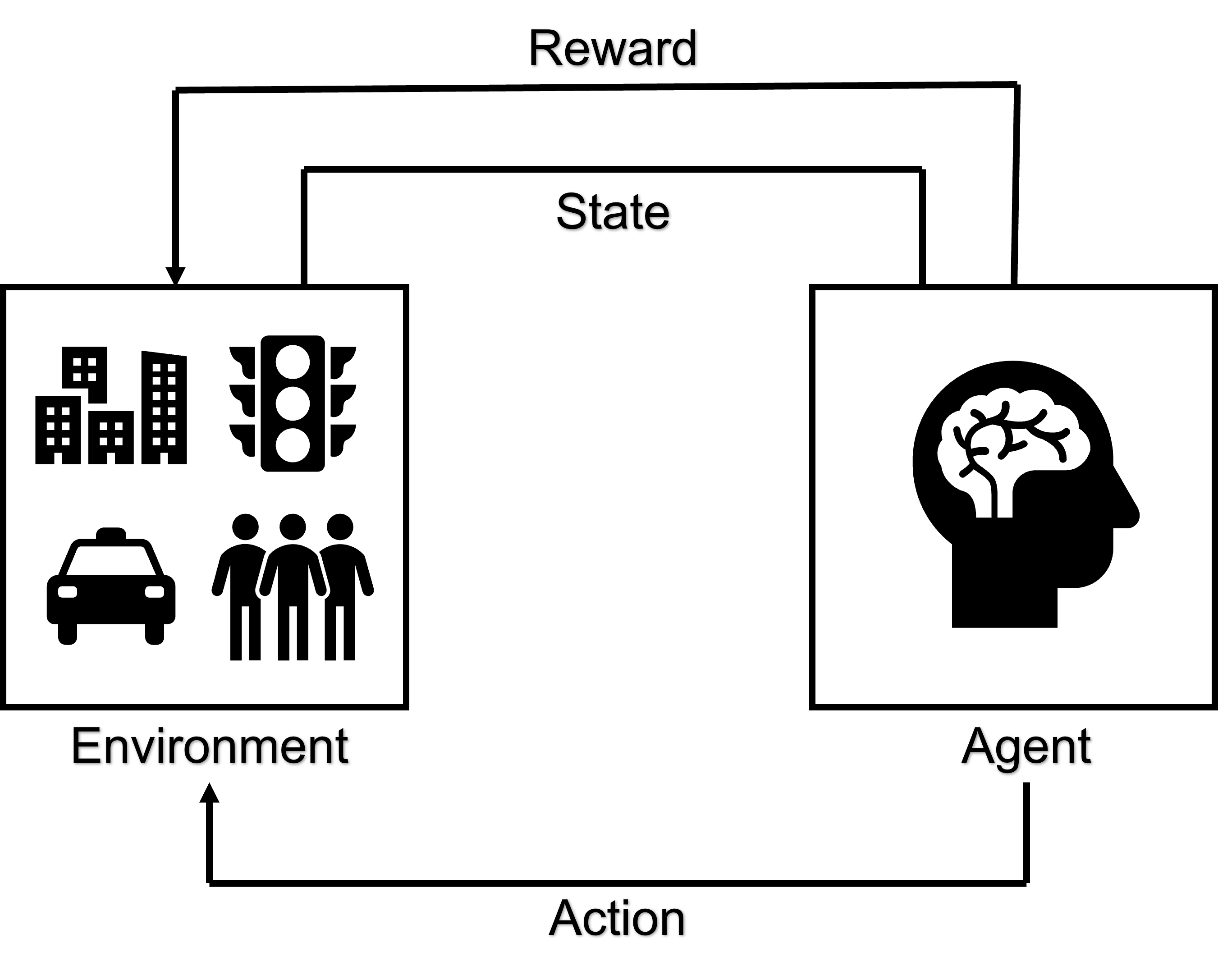}
	\caption{Reinforcement learning control loop}
	\label{fig000}
\end{figure}

The traditional RL method requires detailed information about the controlled object, which hinders the application of RL in many real-life situations. The advance of the model-free RL algorithm broke this barrier and made RL more popular and effective. Among different model-free RL algorithms, Q-learning method\cite{watkins1992q} and State-action-reward-state-action (SARSA)\cite{rummery1994line} are two widely used algorithms. Furthermore, the combination of DL and RL, which is also known as DRL, enables RL to deal with controlled objects with high-dimensional state space. In the high-dimensional and complex system control scenarios, DRL methods with DL network acting as the action decision center, are proved to be more effective than traditional RL algorithms because traditional RL methods with linear function approximation methods cannot effectively handle the large state space and the complex system. Among those DRL methods, Deep Q Network (DQN) and Dueling Double Deep Q Network (Dueling DDQN) are two widely used methods. Nowadays, there are multiple applications of DRL in the ITS. The most popular application is DRL-based traffic signal control\cite{zhang2019cityflow}.

As a data-driven method, RL experiment requires a big amount of data. Nowadays, many RL experiments are conducted with real-world data. Also, traffic simulation software could be alternative when real-world data is inaccessible\cite{huang2022incremental}. SUMO, TESS NG, PTV VISSIM, Paramics, AIMSUN, Transmodeler, Trafficware, Cube Dynasim are all useful software for traffic simulation.

Even though DRL is a powerful tool for traffic control, it still suffers from poor data-efficiency\cite{greguric2020application}. In the nalnal control, DRL requires a long temporal horizon. During the backpropagation of the reward signal, the long sequence may lead to problems of gradient vanishing. Moreover, the reward could be affected by noise\cite{kalweit2019composite}. At the same time, DRL models are less scalable since the decision space is large and the computation cost of the DRL method is tremendous. For real-world application, multi-agent control is preferred. However, the great computation cost requires qualified hardware to support the computation. Also, the interaction of multiple agents and the environment makes the problem more difficult to solve. Even though the multi-agent method is applied, how to ensure the policy provided by each agent can make the whole network globally optimized is also a big problem.

As for vehicle dispatching and navigation, there are multiple works about vehicle dispatching using RL methods, including taxi dispatching and ambulance dispatching. In these works, the traffic network is regarded as non-stationary and stochastic, being composed of different road conditions\cite{mao2018reinforcement}. Generally, there are two main methods for vehicle dispatching, which include route planning and Traffic Assignment Problem (TAP)\cite{qin2021reinforcement}. Route planning, which could also be regarded as route choice, realizes the vehicle dispatching by selecting a given Origin-destination (OD) pair from a set of feasible routes. When applying the route planning method, the action given by the agent is only reviewed after the trip is finished. For TAP, the RL agent needs to make a decision on the next link on the basis of state information, and rewards for each decision are given after every agent environment interaction.
\begin{figure}[!htb]
	\centering
	\includegraphics[width=1\linewidth]{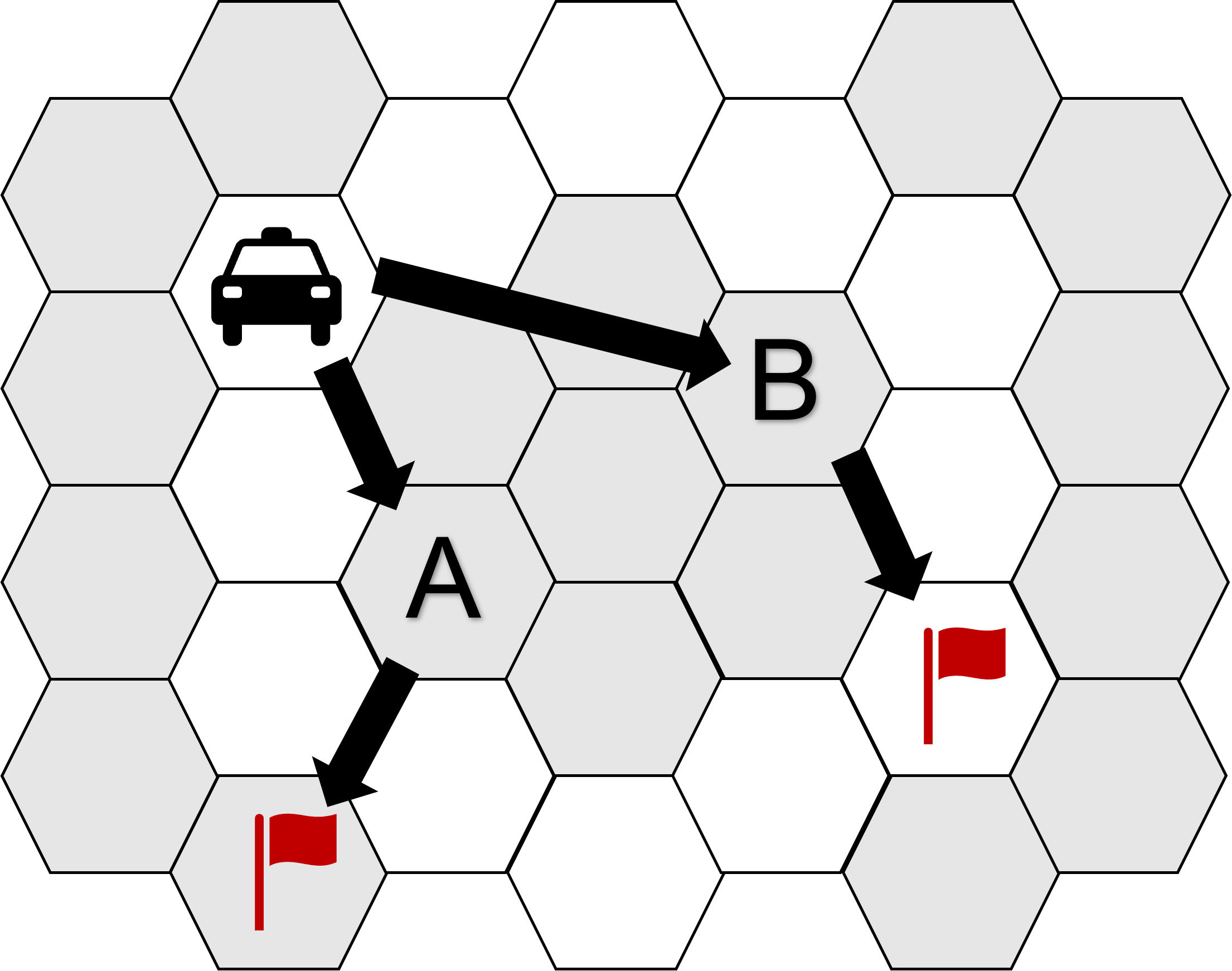}
	\caption{Single vehicle dispatching in a hexagon grid-like network}
	\label{fig001}
\end{figure}

In different scenarios, elements including states of the object and reward given by the environment, are defined differently with different optimization purposes and RL training methods. The majority of RL-based methods for vehicle dispatching utilize two types of information as object states, which include the vehicle state and the system state. Vehicle state could be different information about the vehicle such as the location, travel time, remaining capacity for taxi or battery level for EV\cite{ulmer2020modeling}. As for system information, states could be the location of customers for taxi dispatching systems, locations of different EV charging stations for EV navigation systems or traffic flows in different parts of the traffic network. Among those RL-based methods, value-based RL is the most commonly used method with the optimization purpose of realizing User Equilibrium (UE). Author of \cite{rong2016rich} proposed a value-based value iterative method to solve the MDP, using trip fare as environment reward and direction from which the driver arrives as object states to train the RL agent. Some other researchers utilize not only information from the target vehicle but also contextual information as object states. In \cite{jiao2021real}, local information together with contextual information of the current target is taken by the agent as states information to train an offline CVNet.

Furthermore, in many works, vehicle dispatching or vehicle routing focus on only 1 vehicle using the single agent RL method. In the case of the vehicle fleets, multi-agent RL methods should be applied, controlling multiple vehicles simultaneously. In this situation, the RL agent should be able to realize the global optimization because the action on an object may affect other objects nearby and the optimization of a vehicle may not lead to the greatest reward for the whole system. In \cite{zhou2019multi}, the author proposed a novel multi-agent DRL method for order-vehicle dispatching. Unlike the traditional multi-agent RL method, model agents in the paperwork separately with the guidance from a joint-policy evaluation. The team of Allen proposed an OpenAI Gym based framework for ambulance allocation experiment using DRL methods\cite{allen2021developing}. Liang designed a multi-agent DRL scheduling method for shared EV fleet\cite{liang2020mobility}, which includes EV dispatching to fulfil resource-demand optimization, EV rebalancing and EV charging scheduling from a new angle of power resource.

However, although there are numerous promising designs for vehicle dispatching or allocation with state-of-art model performances, some challenges still exist. First, EVs and autonomous vehicles are becoming a heterogeneous part of the traditional vehicle dispatching problem because both of them are distinct from traditional vehicles. For example, the operation range of EV is limited by the battery level. Routing of EV to a nearby charging station is required when the battery level is low but high enough for the trip to the charging station. Autonomous vehicles may drive in a predefined route and handle emergencies differently compared to human drivers. All these differences are reminding researchers to find out a way that can enable the routing system to deal with a traffic environment that is composed of different kinds of vehicles. Second, works about simulation environments are not sufficient for now. When the real-world data becomes costy\cite{luo2020traffic} to obtain or apply, RL experiments need to rely on traffic simulation environments. Even though many works about simulation environments are proposed, there is still a great need for a solid experiment environment which is similar enough to the real world.


Motivated by these related works, with the purpose of maximizing the usage of different EV charging stations, providing users with more useful information and minimizing the EV queuing and charging time, we decide to design the DRL-based charging navigation system with a suitable simulation environment. The navigation system is an attempt to build an environment for RL experiments using the traffic simulator SUMO. In the traffic area, SUMO is preferred by many researchers for its diverse functions and realistic simulation. Numerous works are implemented using SUMO. For example, SUMO is widely used to simulate different traffic light control schemes. However, the number of works that utilize SUMO to build an RL environment for EV navigation or scheduling is relatively small. Therefore, our EV charging navigation system can provide some ideas for the following proposals about RL-based EV scheduling systems using SUMO.

\section{METHODOLOGY}
For EV navigation system construction, the main purpose is to build an environment for the RL experiment. Some requirements are in need to be fulfilled. First, it should be an environment for EVs, which means that there are multiple EVs running in the network together with some charging stations. Second, the environment should be similar to the real-world situation with different traffic conditions. Third, the system should be able to retrieve vehicle information and execute the deployment. In general, the navigation system together with different EV deployment strategies aim to help EV users get charged with minimum travel time and queuing time.

\subsection{Problem Statement}
In this paper, we are considering an environment in which several EV charging stations are geographically distributed in the traffic network and a minimum distance is ensured between stations. Inside the traffic network, traditional vehicles and EVs are travelling together with random OD pairs but they imitate and follow a driving pattern of real-world vehicles. Moreover, EVs with charging demand will select a nearby charging station with a minimum distance. In this scenario, there will be busy road sections and peak travel times where congestion possibly exists and there will be queuing in charging stations that are near busy road sections at peak travelling time. We consider a single target EV that is driving with a random OD pair and under the guidance of a modern EV charging navigation system. In the navigation system, there is a navigation center obtaining information about the target EV such as real-time longitude, latitude or speed which are sent by sensors on the target EV. At the same time, sensors deployed all over the traffic network send the traffic situation information to the navigation center from different parts of the traffic network. Moreover, charging and queuing conditions could also be known by the navigation center.

In the network, we split the time into a sequence of time slots which could be represented as $T=[t_0, t_1, \cdots, t_l]$ where $l$ is the considered total time length and the size of all time slots are fixed as $\delta t$\cite{liu2020context}. We define the state of the target vehicle at time $t_i$ is $S_i=[s_i^1, s_i^2, \cdots, s_i^k]$ where k refers to the total number of different states. Charging station set is denoted by $C=[c_1, c_2, \cdots, c_m]$ where m is the total number of charging stations, and the action taken at time $t_i$ is represented as $a_i$. Based on the problem background, our purpose is to build an EV charging navigation system that can find out a proper action ai
to minimize the total travelling time for EV drivers given the current states Si of the target EV using the function:
\begin{equation}
a_i=f(S_i)
\end{equation}

\subsection{EV Charging Navigation System Design}
\subsubsection{Definition of the Navigation System}
In this subsection, we will describe our EV charging navigation system design in detail. Definition of target vehicle states, actions and the final evaluation standard for the system performance would be firstly presented. After that, four different dispatching methods for action decisions would be shown including random dispatching, greedy dispatching and RL-based method including DQN and Dueling DDQN.

For target vehicle states, three kinds of information which could be classified into two types, external and intrinsic information, are selected. Intrinsic information we select for target vehicle states is the distance to each EV charging station which could be represented as $D_i=[d_i^1, d_i^2, \cdots, d_i^m]$ where $i$ refers to the time index $t_i$. For external states, what we adopt are road condition parameterized value and charging station usage. Road condition parameterized value reveals the degree of congestion on the way to different charging stations. If there is a serious traffic jam in the chosen route, a big amount of time would be wasted for EV charging. In this study, we choose the number of vehicles on the way to charging stations to represent parameterized road condition value. At time $t_i$, it could be denoted by $N_i=[n_i^1, n_i^2, \cdots, n_i^m]$. What we choose for the other external state is the number of vehicles in charging stations which could be similarly denoted by $Z_i=[z_i^1, z_i^2, \cdots, z_i^m]$ at time $t_i$. Under the definition of states, vehicle state $S_i$ at time $t_i$ becomes $S_i=[D_i, N_i, Z_i]$ and Equation(1) could be further derived as:
\begin{equation}
a_i= f(D_i, N_i, Z_i)
\end{equation}

Actions in EV navigation are the advice and control signals given by the navigation center. In many works about taxi dispatching or navigation using RL methods, the traffic network is interpreted by a composition of rectangle or hexagon grids. Actions in these works are defined to be the destination of a taxi\cite{mao2020dispatch}. In some other works, there is more detailed guidance given by the navigation center. For example, some navigation center provides road-by-road guidance to vehicles instead of the destination\cite{dantzig1959truck}. In this study, the action we are considering is the destination for the target EV. At each time slot, the navigation center observes intrinsic and external states on the basis of which the center generates a control signal, an advised charging station, to the target EV. The cycle repeats until the target vehicle stops for charging.

For the final evaluation standard, the main purpose of the EV navigation system is to minimize the travelling time for EV drivers. Therefore, we estimate the total travelling time from the target vehicle departure to the final EV charging.

\begin{figure}[!htb]
	\centering
	\includegraphics[width=1\linewidth]{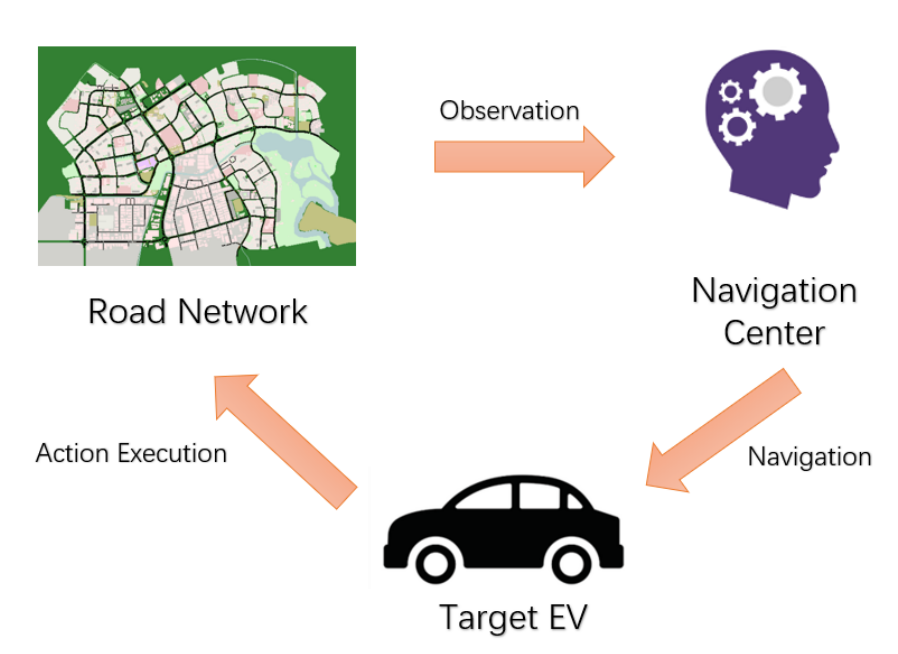}
	\caption{Navigation system operation cycle}
	\label{fig003}
\end{figure}

\subsubsection{Action Determination Methods}
Four methods are chosen for action determination including random dispatching, greedy dispatching, DQN and Dueling DQN. The random dispatching scheme provides a random selection of destination charging station from the charging station pool. The answer is given randomly without any consideration of current states and the random action is given only at the time when the target vehicle enters the traffic network because a series of random guidance will lead to meaningless driving inside the network. The action determination at time $t_i$ for the random dispatching scheme can be represented as:

\begin{equation}
a_i=Random(C)
\end{equation}

As for the greedy dispatching scheme, the navigation center only guides the target vehicle to the closest charging station on the basis of the current target EV location. This kind of control scheme regards the journey with the smallest distance as the most rewarding solution but ignores the road condition and availability of each charging station. Action determination at time $t_i$ for the greedy dispatching scheme can be represented as:
\begin{equation}
a_i = \argmin_{a} D(i)
\end{equation}

The first two navigation schemes only take one kind of information or even no information into consideration, neglecting the bigger picture of the road network. Even though in some scenarios, random dispatching or greedy dispatching schemes work, they cannot handle most of real-world situations. For example, when the closest charging station is in an area where the traffic flow is huge and a big number of EV drivers are queuing for charging, avoiding the busy road section for another charging station is a better choice. A better navigation system which could also be regarded as a smart agent is required to solve this situation by taking more information into consideration. In recent years, researchers are adopting RL methods for vehicle dispatching especially DRL methods DQN and Dueling DQN\cite{he2020spatio}\cite{oda2018movi}.

Traditional RL methods are designed on the basis of the MDP framework, where the agent follows a policy $\pi(s)$ to determine actions. Sufficient training enables the RL agent to find out the optimized policy which leads to the greatest short-term or long-term reward $R$. Q-learning is an off-policy temporal-difference learning approach. It can obtain the maximum long-term discount reward $Q(S,A)$, where $A$ refers to the action space. The policy update of the Q-learning function is expressed as:
\begin{equation}
Q^{\ast}(s,a) \gets Q(s,a) + \alpha \left[r+\gamma \max_{a^{\ast}}Q(s,a^{\ast})-Q(s,a) \right]
\end{equation}
where $\alpha$ and $\gamma$ refer to the learning rate and the discount factor respectively. When the state space and action space become too large, which is common in real-world scenarios, the pure mathematical computation cannot handle the large state and action space optimized policy solving. With the adoption of the DL network approximating the Q function, DQN is able to deal with large state space while ensuring the model performance.

During the training, we deploy the target EV into the simulation environment. DQN agent observes the states of the target at every time slot, determining actions for the target. After the action execution, the environment reacts with immediate reward. In this study, we define all the immediate rewards to be 0 until the target EV stops and starts to charge. The final reward is defined as:
\begin{equation}
R=\frac{7200}{T_{travel}}
\end{equation}
where $T_{travel}$ refers to the total travelling time of the target vehicle from departure to the final EV charging. A simulation cycle is one day in the simulation environment and this cycle repeats to enhance the stability of DQN training. Table 1 is an example agent-environment interaction table in which the agent cannot get feedback of reward until the target vehicle stops at $t_3$.
\begin{table}[!htbp]
\centering
\caption{An Example Simulation Cycle}
\resizebox{0.3\textwidth}{!}{
\begin{tabular}{cccc}
\hline
 & Action & Reward & Done\\
\hline
$t_1$ & $a_1$ & 0 & False\\
$t_2$ & $a_1$ & 0 & False\\
$t_3$ & $a_1$ & R & True\\
\hline
\end{tabular}
}
\label{table1}
\end{table}

To guarantee the training stability, we use a prioritized experience replay method which buffers some simulation cycles and trains the agent with a random new policy $\pi(s)$ instead of the well-trained one\cite{43}. By applying the experience replay, the DQN agent can avoid generating the same answer which may lead to local optimization but not global optimization in certain situations, and becomes able to explore new possibilities for action determination. Also, a target network\cite{44} is utilized to ensure the training stability. The target network can keep the Q-value to be constant for a while, reducing the correlation between the current Q-value and the target Q-value. At the same time, when updating Q-network weights, the Mean-squared Error loss function $L(\theta)$ is used. $L(\theta)$ calculates the difference between the predicted Q-values and the target Q-values.
\begin{equation}
L(\theta)=E\left[((r+\gamma \max_{a^{\ast}}Q(s,a^{\ast};\theta^{'})-Q(s,a;\theta)))^2\right]
\end{equation}
where $\theta$ and $\theta^{'}$ are weights of behavior network and target network respectively. Algorithm 1 presents the pseudocode of Deep Q-learning with prioritized experience replay. At the beginning, we initialize the simulation environment and the Deep Q-learning agent. Memory $D$ which storages simulation information including the last-step state $s_{i-1}$, last-step action $a_{i-1}$, last-step immediate reward $r_{i-1}$ and current step state $s_i$ at time slot $t_i$, is set to be empty. Capacity $N$ represents the size of memory. Weights $\theta$ and $\theta^{'}$ of the behavior Q-network and the target Q-network are initialized with random values respectively. Then the DQN is trained with specified max episodes. In each episode, the simulator simulates a traffic network in a day. The target EV is deployed at a random time step with a random OD pair into the traffic network. Following steps of the pseudocode focus only on the target EV. Other vehicles, traditional vehicle and other EVs travels in the traffic network. At each time slot, the agent observes the environment and generates action on the basis of action determination policy but it is possible that the agent will choose a random action for training stability and new solution exploration. The possibility of the random action is $\xi$. After that, the target EV executes the action and the environment gives feedback to the DQN agent. Information about the EV-environment interaction of this step will be memorized by $\mathcal{D}$ and the DQN agent replays some former cycles to understand the situation better and optimize the action decision policy. In the end, the Q-network updates.

\begin{algorithm}{
		\textsl{}\setstretch{1}
		\caption{Deep Q-learning with Experience Replay}
		\LinesNumbered
		\emph {Initialize: memory $\mathcal{D}=\emptyset$ and capacity $N$}\;
		\emph {Initialize: $Q$-network weights $\theta$ with random values}\;
		\emph {Initialize: target $Q$-network weights $\theta^{\prime}$ with random values}\;
		\For{$episode=1$ \KwTo $max-episodes$}{
			\emph {Initialize state $S$ with initial state $S_{0}$}\;
			\For{$\text{step } i=1$ \KwTo $max-steps$}{
				\emph {observate current state $S_{i}$}\;
				\emph {with probability $\xi$ select a random action $a_{t}$}\;
				\emph {otherwise select $a_{t}=\mathop{argmax}\limits_{a}Q\left( s,a;\theta \right)$}\;
				\emph {execute action $a_{t}$ and obtain reward $r_{t-1}$}\;
				\emph {store tuple $\left( s_{t-1}, a_{t-1}, r_{t-1}, s_{t} \right)$ into $\mathcal{D}$}\;
				\emph {sample a random minibatch of memory tuple from $\mathcal{D}$}\;
				\emph {set $$
					y_{i} =
					\begin{cases}
					r_{i} & Terminate@$i+1$ \\
					r_{i}+ \gamma \mathop{max}\limits_{a^*}Q\left( s,a^*;\theta^{\prime} \right)  & otherwise \\
					\end{cases}
					$$}\
				\emph {perform a gradient descent}\;
			}
		}
	}

	\label{Algorithm1}
\end{algorithm}

Even though DQN is proved to be powerful for EV navigation, it suffers from a problem of Q-value overestimation. In DQN, the predicted Q-value is calculated as $y_t^{DQN}=R_{t+1}+\gamma \max_{a}Q(S_{t+1},a;\theta_t^{-})$ which always selects the action that leads to the maximum predicted Q-value. To solve this problem, DDQN is designed. Different from DQN, DDQN uses an action network to decide actions. The estimated Q-value in DDQN is represented as:
\begin{equation}
y_t^{DQN}=R_{t+1}+\gamma Q(S_{t+1}, \argmax_{a} Q(S_{t+1},a;\theta_t);\theta_t^{'})
\end{equation}

Further, in DQN network or DDQN network, a DL network calculates the current Q-value. The final output is generated by a fully connected layer. However, in the Dueling DDQN network, the Q function is separated into two parts which are the state function and the advantage function:

\begin{equation}
Q(s,a;\theta,\alpha,\beta)=V(s,\theta,\beta)+A(s,a;\theta,\alpha)
\end{equation}

where $V(s,\theta,\beta)$ is the state function and $A(s,a;\theta,\alpha)$ is the advantage function. $\theta$ is a parameter in the convolutional layer, and $\alpha$ and $\beta$ are fully connected layer parameters of the state function and the advantage function respectively. This enables the network to have a better estimation of the Q-value. Similar as algorithm 1, the pseudocode of Dueling DDQN is shown in algorithm 2.

\begin{figure}[!htb]
	\centering
	\includegraphics[width=0.7\linewidth]{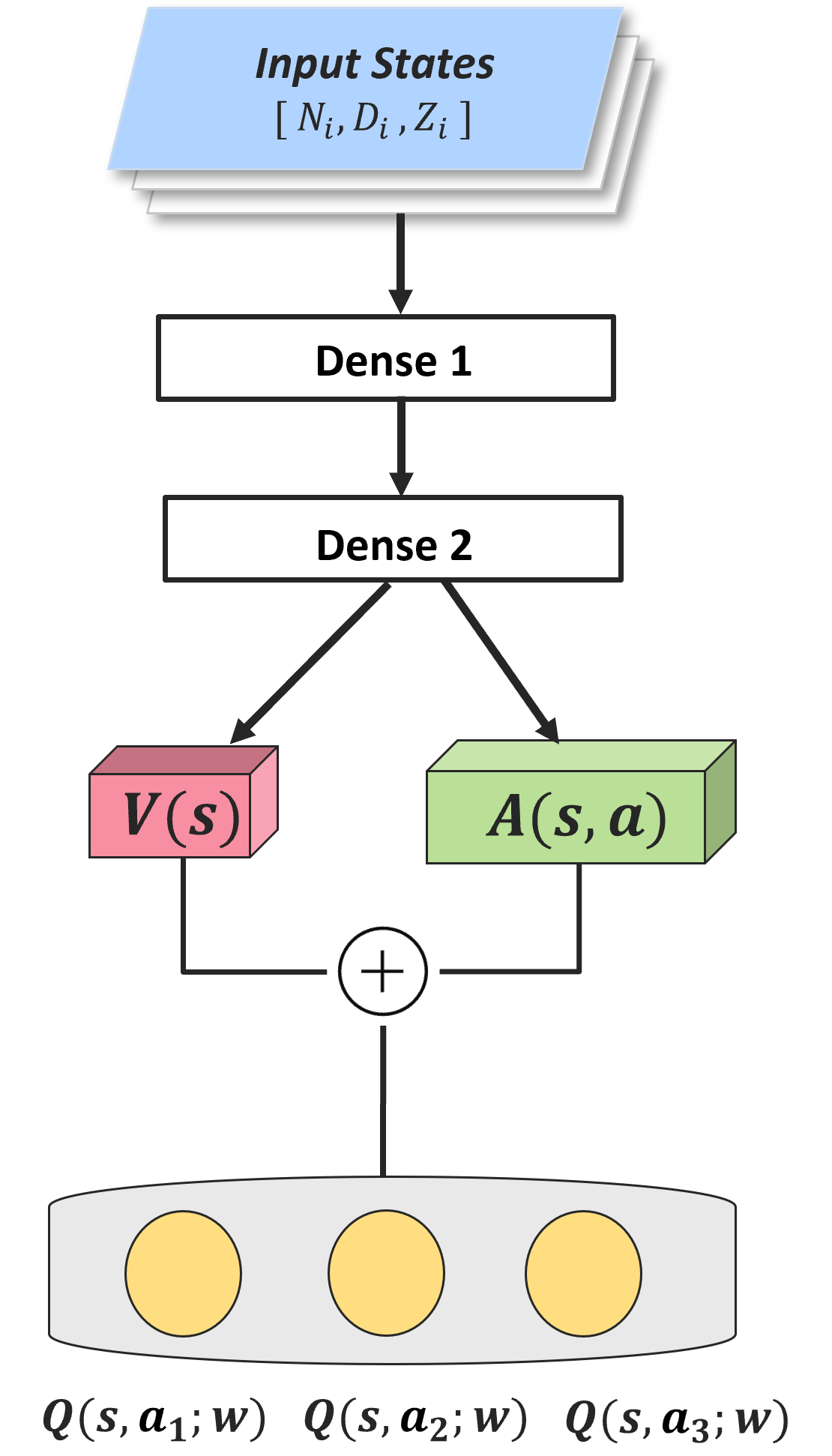}
	\caption{Architecutre of Dueling DDQN in EV charging dispatching problem}
	\label{fig004}
\end{figure}

\begin{algorithm}{
		\emph {Initialize: memory $\mathcal{D}=\emptyset$ and capacity $N$}\;
		\emph {Initialize: $Q$-network weights $\theta$ with random values}\;
		\emph {Initialize: target $Q$-network weights $\theta$ with random values}\;
		\For{$episode=1$ \KwTo $max-episodes$}{
			\emph {Initialize state $S$ with initial state $S_{0}$}\;
			\For{$\text{step } i=1$ \KwTo $max-steps$}{
				\emph {observate current state $S_{i}$}\;
				\emph {with probability $\xi$ select a random action $a_{t}$}\;
				\emph {otherwise select $a_{t}=\mathop{argmax}\limits_{a}Q\left( s,a;\theta \right)$}\;
				\emph {execute action $a_{t}$ and obtain reward $r_{t-1}$}\;
				\emph {store tuple $\left( s_{t-1}, a_{t-1}, r_{t-1}, s_{t} \right)$ into $\mathcal{D}$}\;
				\emph {sample a random minibatch of memory tuple from $\mathcal{D}$}\;
				\emph {set $$
					y_{i} =
					\begin{cases}
					r_{i}   \qquad\qquad\text{if episode terminates at step } $i+1$ \\
					r_{i}+ \gamma Q\left( s^{\prime}, a^{*}\left( s^{\prime},\theta \right);\theta^{-} \right)  \qquad\qquad otherwise \\
					\end{cases}
					$$}\
				\emph {perform a gradient descent}\;
			}
		}
	}
	\caption{Dueling Double Deep Q-learning with Experience Replay}
	\label{Algorithm2}
\end{algorithm}

In conclusion, the EV navigation system is an environment in which a smart agent guides the target vehicle to a charging station with minimum travelling time. A number of traditional vehicles and EVs are travelling in the traffic network together with the target EV. The smart agent observes information from the target and the environment and generates the control signal. Four methods can be adopted by the agents including random dispatching, greedy dispatching, DQN and Dueling DDQN. This system can become a platform for future DRL experiments.

\section{TEST AND EXPERIMENTS}

\subsection{Overview}
In this part, the experiment of the EV navigation system will be described in detail. The following part of this section is organized as follows. We will first describe the simulation environment which is the foundation of the whole experiment. Second, basic environment settings will be presented. Third, we will present our experiment settings and evaluation standard. In the end, the experiment results would be shown.

\subsection{Simulation Environment and Basic Settings}
SUMO is an open source, microscopic and multimodal traffic simulation software. It can simulate a complex traffic network with all the vehicle in the network travelling with their own defined trip. The simple user interface allows users to master the use of this software easily and the Traci interface enables users to realize a more complex traffic simulation with Python. SUMO starts to be widely used from 2001 and nowadays, a big number of works in the traffic area deploy their experiment on SUMO. In this study, the experiment is conducted on the basis of SUMO.

First of all, the traffic network we are simulating is a network in Jurong West, Singapore. Jurong West is a residential town which locates in the west of Singapore. The traffic network of Jurong West is shown in Fig.5.

\begin{figure}[!htb]
	\centering
	\includegraphics[width=0.9\linewidth]{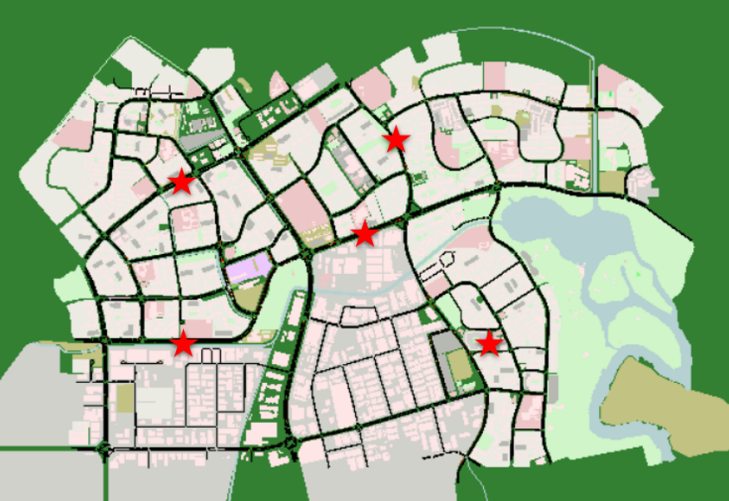}
	\caption{Traffic network in Jurong West}
	\label{fig005}
\end{figure}
Inside the traffic network, 5 charging stations are geographically distributed and they are marked by stars. To realize better training of the RL agent, 4 of the 5 charging stations are artificially added to the network and a minimum distance is kept between each of the 2 stations. The power of each station is 100KW which simulates the real-world charging station

To better simulate the real-world traffic network, we define all the vehicles in the traffic network to follow the driving pattern of real-world drivers. There would be busy road sections, peak travelling time and queuing for charging in the simulation.

\begin{figure}[!htb]
	\centering
	\includegraphics[width=0.9\linewidth]{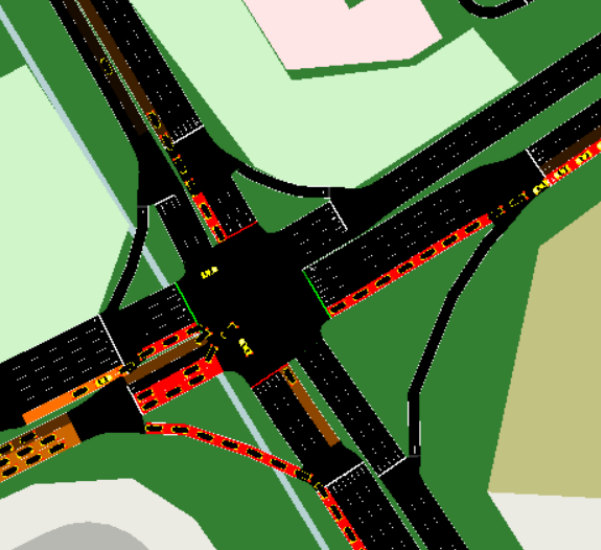}
	\caption{Busy road section}
	\label{fig006}
\end{figure}

\begin{table*}[ht]
	\centering
	\caption{Travelling time of different algorithms in EV Navigation problem}

		\begin{tabular}{ccccc}
			\hline
			Vehicle Number & Random Dispatching & Greedy Dispatching & DQN & Dueling DDQN\\
			\hline
			200 & 942 & 750 & 545 & 632\\
			300 & 1377 & 1158 & 633 & 587\\
			400 & 1955 & 1648 & 1387 & 843\\
			\hline
		\end{tabular}
	
	\label{table2}
\end{table*}

During the experiment, each episode simulates for a simulation day and the total number of episodes is 50. The time in the experiment refers to the simulation time in the SUMO environment which is accelerated compared to real time. Based on the basic definition, the experiment can be summarized as follows. An episode of simulation begins at 00:00 in the morning. A target vehicle departs at a random simulation time and seeks a charging station. During the simulation, the target vehicle is guided by a smart agent. The target EV stops when it begins to charge and the episode does not end until 23:59. This cycle repeats 50 times.

To evaluate the performance of each action determination scheme, we use the total travel time in the simulation environment of the target EV as our evaluation standard.

\begin{figure}[!htb]
	\centering
	\includegraphics[width=1\linewidth]{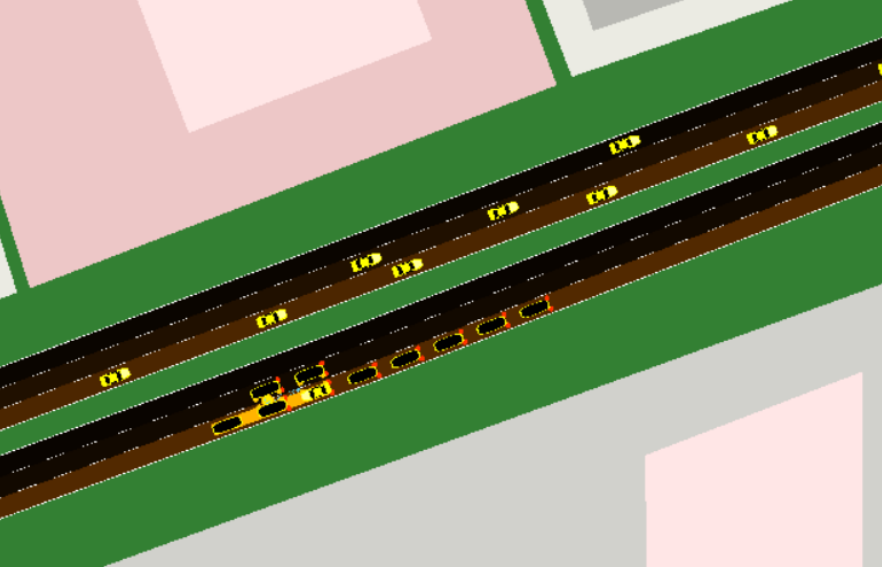}
	\caption{EV queuing for charging}
	\label{fig007}
\end{figure}

For the training of DQN and Dueling DDQN network, both of the networks are optimized by Adam optimizer. Learning rates for both of the networks are set to be 0.01 and the training batch size is 32. All of the experiments are conducted on a GPU RTX3060. The parameters details are listed in the following table.

\begin{table}[ht]\centering
        \caption{The tuning parameters of deep reinforcement learning model}\label{tab:list data interface}
         \begin{tabular*}{\hsize}{@{}@{\extracolsep{\fill}}lc@{}}\toprule
         
        Parameters  &  Value \\\midrule
        First hidden layer size& 512 \\
        Second hidden layer size& 256 \\
        Learning rate & 0.01 \\
        Training batch size & 32 \\
        Target network update frequency & 8000 \\
        Greedy epsilon & 1 $\rightarrow$ 0.1 \\
        Buffer size & 10000 \\
        Max episode & 50 \\
        \bottomrule
        \end{tabular*}
        \end{table}

\subsection{Experiment Results}
To better evaluate the system performance, we deploy 3 different numbers of EVs into the simulation environment for each experiment while the number of traditional vehicles is fixed. All of the vehicle navigation schemes are evaluated in the environment with 200, 300 and 400 EVs respectively. Experiment results are shown in Table 4.2 where values of each dispatching scheme refer to the average simulation time of all the 50 episodes in the artificial environment.

\begin{figure}[!htb]
	\centering
	\includegraphics[width=1.1\linewidth]{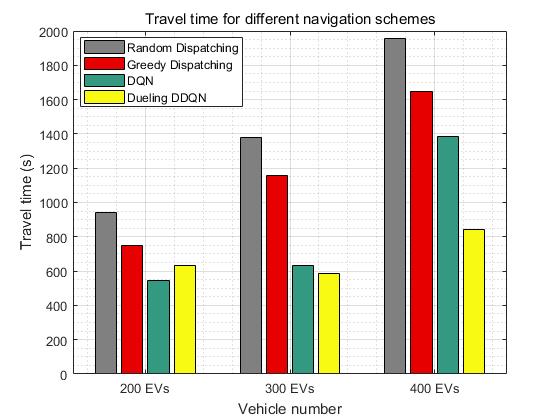}
	\caption{Travel time for different navigation schemes}
	\label{fig008}
\end{figure}

\subsection{Discussion}
According to Table 4.2, the DQN method and Dueling DDQN method have a better performance in EV navigation with less total travelling time since RL-based methods take more information into consideration and they learn a better solution in different road conditions through numerous attempts. Greedy dispatching which requires a longer time for station navigation compared to RL-based methods performs better than random charging station selection since the random selection may guide the vehicle to a distant station which wastes a large amount of time. Although Dueling DDQN requires a longer navigation time in the 200-EV scenario, when the number of EVs becomes bigger and the road condition becomes more and more complex, Dueling DDQN starts to outperform the DQN model. In the 400-EV scenario, Dueling DDQN takes only 843 seconds of simulation time which is smaller than that of DQN by 39$\%$. Both the DQN and Dueling DDQN are successfully deployed and have achieved a good performance in EV navigation, which means that the system is successfully built for DRL experiments.

\section{CONCLUSIONS}
In this study, we build a SUMO-based EV charging navigation system with 4 different navigation schemes in use which includes random dispatching, greedy dispatching, DQN and Dueling DQN. We apply our system in the SUMO simulated Jurong West traffic network to guide a single target EV to a charging station. The purpose of the navigation is to minimize the total travelling time from the vehicle departure to battery charging. In the experiment, DQN and Dueling DDQN are proved to be effective in the SUMO environment which means that the system we build is a useful environment for future RL experiments. In the future, the EV navigation environment can be enriched by more real-world situations including different weather, traffic jam, accidents and pedestrian. We will try to combine the traffic forecasting model with the navigation system to improve the ability of EV navigation and adopt multi-agent DRL models to realize multiple vehicle navigation and global optimization for the whole traffic network.





\section*{ACKNOWLEDGMENT}

This study is supported under the RIE2020 Industry Alignment Fund – Industry Collaboration Projects (IAF-ICP) Funding Initiative, as well as cash and in-kind contribution from the industry partner(s).

\ifCLASSOPTIONcaptionsoff
  \newpage
\fi


\singlespacing
\bibliographystyle{IEEEtran}
\bibliography{autosam}


\begin{IEEEbiography}[{\includegraphics[width=1in,height=1.25in,clip,keepaspectratio]{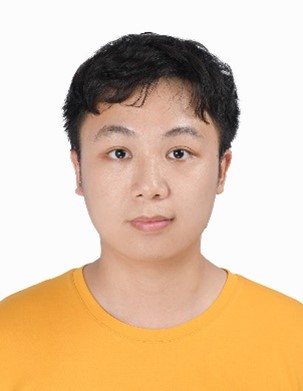}}]{Yaofeng Song}
	received the bachelor degree from the school of Automation Science and Engineering in South China University of Technology. Currently he is a Msc student in the school of Electrical and Electronic Engineering in Nanyang Technological University, Singapore. His research interests invlove deep learning based traffic forecasting.
\end{IEEEbiography}

\begin{IEEEbiography}[{\includegraphics[width=1in,height=1.25in,clip,keepaspectratio]{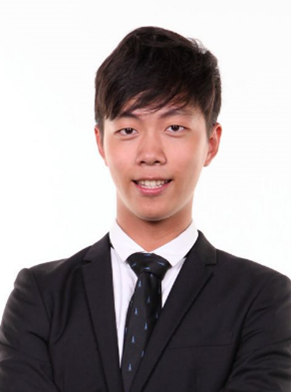}}]{Han Zhao}
	received the Bachelor degree in Electrical and Electronic Engineering from Nanyang Technological University, Singapore in 2018. He is currently working toward the Ph.D degree in Electrical and Electronic Engineering in Nanyang Technological University, Singapore. His research interests include intelligent transportation system (ITS), short-term traffic flow prediction and graph neural networks.
\end{IEEEbiography}

\begin{IEEEbiography}[{\includegraphics[width=1in,height=1.25in,clip,keepaspectratio]{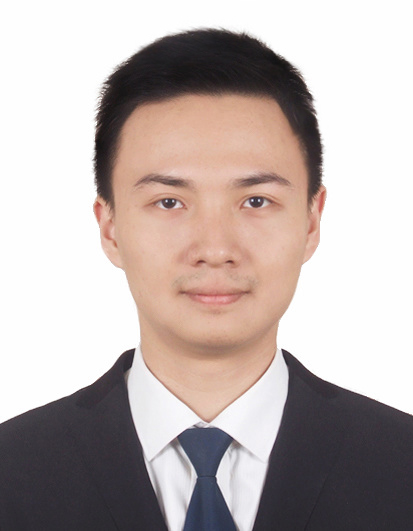}}]{Ruikang Luo}
    received the B.E. degree from the School of Electrical and Electronic Engineering, Nanyang Technological University, Singapore. He is currently currently pursuing the Ph.D. degree with the School of Electrical and Electronic Engineering, Nanyang Technological University, Singapore. His research interests include long-term traffic forecasting based on spatiotemporal data and artificial intelligence.
\end{IEEEbiography}

\begin{IEEEbiography}[{\includegraphics[width=1in,height=1.25in,clip,keepaspectratio]{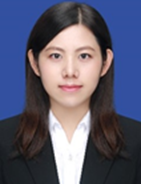}}]{Liping Huang}
	Huang Liping obtained her Ph. D, and Master of Computer Science from Jilin University in 2018 and 2014, respectively. She has been working as a research fellow at Nanyang Technological University since 2019 June. Dr. Huang’s research interests include spatial and temporal data mining, mobility data pattern recognition, time series prediction, machine learning, and job shop scheduling. In the aforementioned areas, she has more than twenty publications and serves as the reviewer of multiple journals, such as IEEE T-Big Data, IEEE T-ETCI, et al.
\end{IEEEbiography}

\begin{IEEEbiography}[{\includegraphics[width=1in,height=1.25in,clip,keepaspectratio]{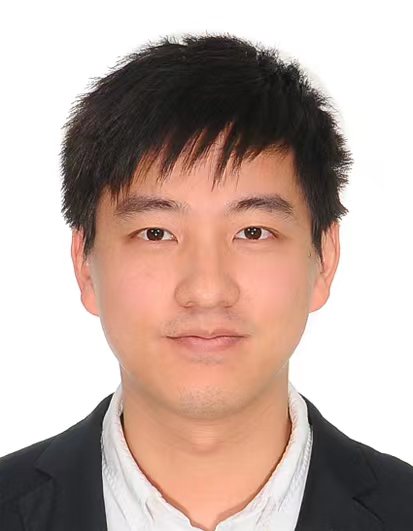}}]{Yicheng Zhang}
	Yicheng Zhang received the Bachelor of Engineering in Automation from Hefei University of Technology in 2011, the Master of Engineering degree in Pattern Recognition and Intelligent Systems from University of Science and Technology of China in 2014, and the PhD degree in Electrical and Electronic Engineering from Nanyang Technological University, Singapore in 2019. He is currently a research scientist at the Institute for Infocomm Research (I2R) in the Agency for Science, Technology and Research, Singapore (A*STAR). Before joining I2R, he was a research associate affiliated with Rolls-Royce @ NTU Corp Lab. He has participated in many industrial and research projects funded by National Research Foundation Singapore, A*STAR, Land Transport Authority, and Civil Aviation Authority of Singapore. He published more than 70 research papers in journals and peer-reviewed conferences. He received the IEEE Intelligent Transportation Systems Society (ITSS) Young Professionals Travelling Scholarship in 2019 during IEEE ITSC, and as a team member, received Singapore Public Sector Transformation Award in 2020.
\end{IEEEbiography}

\begin{IEEEbiography}[{\includegraphics[width=1in,height=1.25in,clip,keepaspectratio]{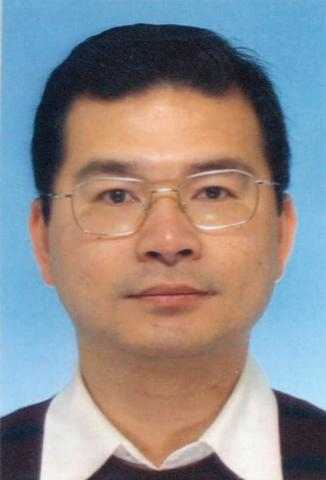}}]{Rong Su}
	received the M.A.Sc. and Ph.D. degrees both	in electrical engineering from the University of Toronto, Toronto, Canada, in 2000 and 2004 respectively.	He is affiliated with the School of Electrical and Electronic Engineering, Nanyang Technological University, Singapore. His research interests include modelling, fault diagnosis and supervisory control of discrete-event dynamic systems. Dr. Su has been a member of IFAC technical committee on discrete event and hybrid systems (TC 1.3) since 2005.
\end{IEEEbiography}

\end{document}